\documentclass[osajnl,twocolumn,showpacs,superscriptaddress,11pt]{revtex4-1} 
\usepackage{amsmath,amssymb,graphicx}
\begin{document}

\title{Frequency stability of a wavelength meter and application to laser frequency stabilization}

\author{Khaldoun Saleh}
\author{Jacques Millo}
\author{Alexandre Didier}
\author{Yann Kersal\'e}
\author{Cl\'ement Lacro\^ute}\email{Corresponding author: clement.lacroute@femto-st.fr}
\affiliation{FEMTO-ST Institute, UMR 6174 : CNRS/ENSMM/UFC/UTBM/UBFC , Time and Frequency Dpt., 26 ch. de l'Epitaphe, 25030 Besan\c{c}on Cedex, France}

\begin{abstract}
Interferometric wavelength meters have attained frequency resolutions down to the MHz range. In particular, Fizeau interferometers, which have no moving parts, are becoming a popular tool for laser characterization and stabilization. In this article, we characterize such a wavelength meter using an ultra-stable laser in terms of relative frequency instability $\sigma_y(\tau)$ and demonstrate that it can achieve a short-term instability $\sigma_y(1s) \approx 2{\times}10^{-10}$ and a frequency drift of order $10$ MHz/day. We use this apparatus to demonstrate frequency control of a near-infrared laser, where a frequency instability below $3{\times}10^{-10}$ from 1 s to 2000 s is achieved. Such performance is for example adequate for ions trapping and atoms cooling experiments.
\end{abstract}


\maketitle 

\section{Introduction}

Laser frequency stabilization is an ubiquitous tool used in several domains such as spectroscopy, laser cooling of atoms, ions or molecules, frequency metrology, atomic frequency standards and sensors or quantum information processing. Whereas frequency metrology and optical atomic clocks require ultra-stable lasers with relative frequency instabilities $\sigma_y$ below $10^{-15}$\cite{Jiang2011}, applications such as atom cooling can be performed with short-term instabilities of order $10^{-10}$ or below, which is sufficient to ensure that the laser frequency is stabilized within the natural linewidth of the involved transitions. This is usually achieved using an atomic reference in a saturated absorption setup, using a neutral atom gas cell \cite{Rovera1994} or ions in a discharge lamp \cite{Streed2008, Lee2014}. When no atomic line can be used, an optical resonator is preferred. For example, Fabry-Perot (FP) cavities or, more recently, optical micro-resonators \cite{Matsko2006, Ilchenko2006} are good candidates for such applications.

On the other hand, these cavities often exhibit high frequency drifts and need to be in turn stabilized to atomic references \cite{Bohlouli-Zanjani2006, McLoughlin2011} and act as so-called "transfer cavities", or to be frequently compared to an atomic line. Recently, interferometric wavelength meters have been demonstrated to be a valid alternative, and have been used both as characterization devices and as frequency stabilization tools \cite{Kobtsev2007, McLoughlin2011, Pyka2014}. Commercial wavelength meters have been recently used in atomic spectroscopy \cite{Kobtsev2007} and in ion trapping and cooling experiments \cite{Pyka2014, McLoughlin2011}. 

Such apparatus have been partly characterized in terms of long-term drift \cite{Kobtsev2007} and frequency stability \cite{Pyka2014}. Nevertheless, to our knowledge, no prior full characterization of both wavelength meters frequency instability and frequency-locking performance have been performed. In this article, we first characterize a commercial wavelength meter in terms of fractional frequency istability $\sigma_y(\tau)$ using the Allan deviation. We show that its resolution limit is well below the specified 10 MHz resolution and that its frequency drift can be reduced down to about $3 {\times} 10^{-13} \ \tau$ (about 10 MHz/day) in a well-controlled environment. Afterwards, we perform a frequency-locking of a telecom laser by the wavelength meter and we measure its frequency stability by comparison with an ultra-stable laser. We finally show that the locking performance of such device is good enough to be used for ion trapping and atom cooling experiments.


\section{Wavelength meter characterization}

\subsection{Wavelength meter}

The wavelength meter we are characterizing and using in this article is a Standard WS7 wavelength meter from HighFinesse. This wavelength meter can perform wavelength measurements for CW and pulsed lasers in the 350 - 1120 nm range. It consists of an 8 L fiber-coupled optical unit based on five Fizeau interferometers that create interference patterns. These patterns are detected by two CCD arrays and transmitted to a software that calculates the wavelength of the laser under test in comparison with an internal reference (a built-in Neon lamp is used to calibrate the device). In addition, a multichannel fiber-switch unit can be used to  measure up to eight lasers and an optional PID regulator analog output can be used to perform a frequency-lock for all these lasers. We plan to use the 4-channel version of the switch to simultaneously measure and stabilize the cooling and ionization lasers at 369.5 nm and 398 nm.


\subsection{Frequency instability characterization setup}

\begin{figure}%
\centering
\includegraphics[width=\columnwidth]{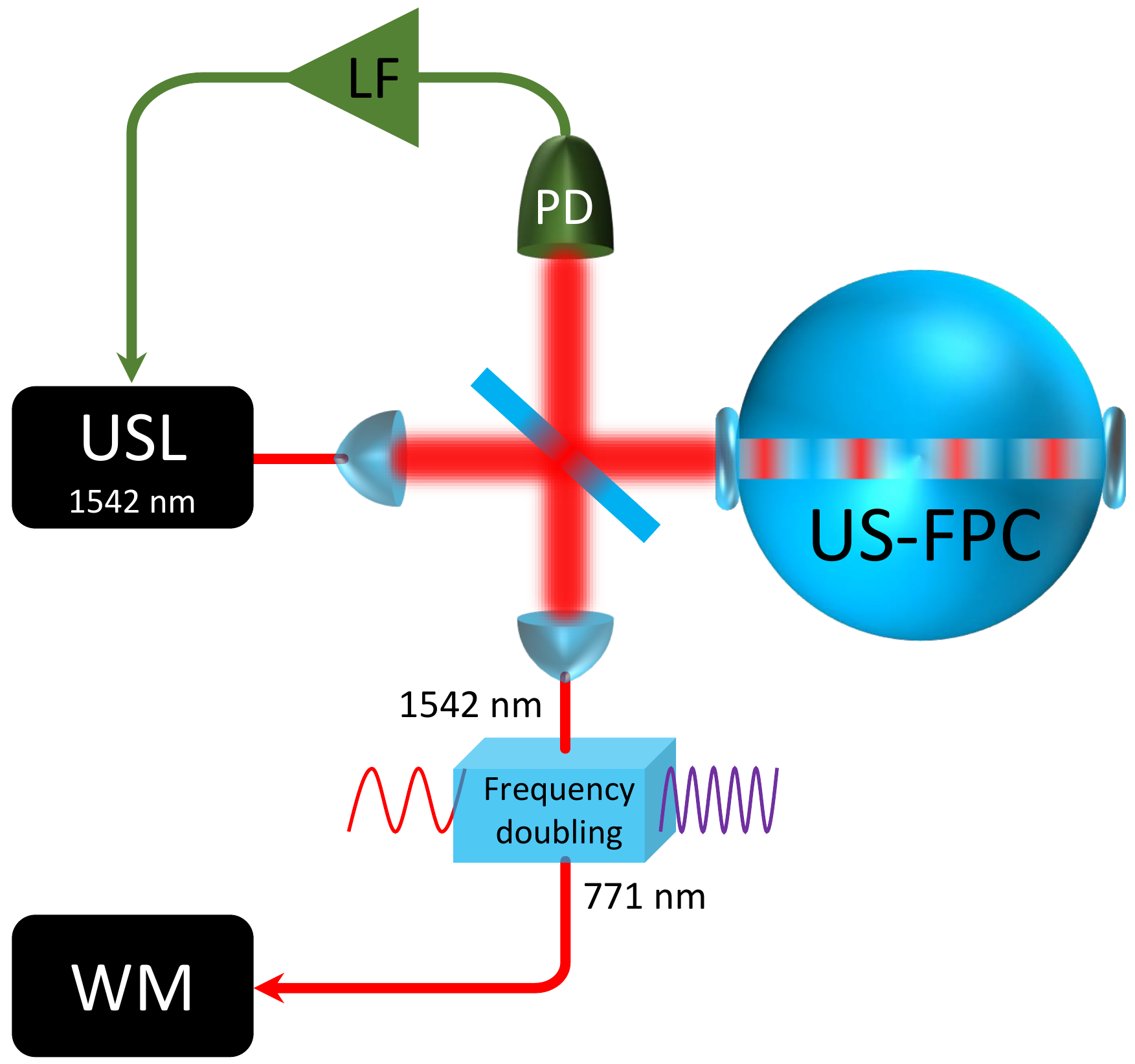}%
\caption{Wavelength meter characterization bench: a 1542 nm laser (USL) is stabilized to an ultra-stable FP cavity (US-FPC) and frequency-doubled using a nonlinear crystal. The resulting signal is fiber-coupled to the wavelength meter (WM). The standard Allan deviation of the optical frequency measured and recorded by the wavelength meter is computed in order to estimate the wavelength meter relative frequency instability. PD: photodiode; LF: loop filter.}.%
\label{fig:setup}%
\end{figure}

In order to characterize the abovementioned wavelength meter in terms of relative frequency stability, we have used the experimental setup depicted in Fig. \ref{fig:setup}. In this setup, a $1542~\mathrm{nm}$ telecom laser (a Koheras AdjustiK$^{\mathrm{TM}}$ E15 from NKT Photonics) is stabilized to an ultra-stable FP cavity featuring a theoretical relative frequency instability of $8{\times}10^{-16}$ at 1s. The stabilized 1542 nm laser was used as an optical frequency reference, as it features a relative frequency instability in the order of $2{\times}10^{-15}$ at 1s \cite{Didier2015}. Afterwards, a beam splitter was used to extract a fraction of the ultra-stable laser beam ($\approx$20 mW) that was then sent to a frequency doubling crystal. Indeed, since our wavelength meter only covers the 350-1120 nm range, we had to multiply the ultra-stable laser frequency. This was done using a nonlinear frequency doubling crystal (a periodically poled lithium niobate (PPLN)) \cite{Taverner1998}. Although the doubling efficiency was relatively low (we obtain 0.7 $\mu$W at 771 nm for a CW injected signal of 20 mW at 1542 nm), the detection threshold of the wavelength meter is on the other hand below $1 \mu$W (0.066 $\mu$W for an exposure time of 150 ms). We therefore always obtained sufficient power at 771 nm to perform our measurements. In the following, we assume that the second-harmonic generation module does not degrade the ultra-stable laser relative frequency instability \cite{Stenger2002, Yeaton-Massey2012}.

\subsection{Frequency instability measurements}

\begin{figure}%
\centering
\includegraphics[width=\columnwidth]{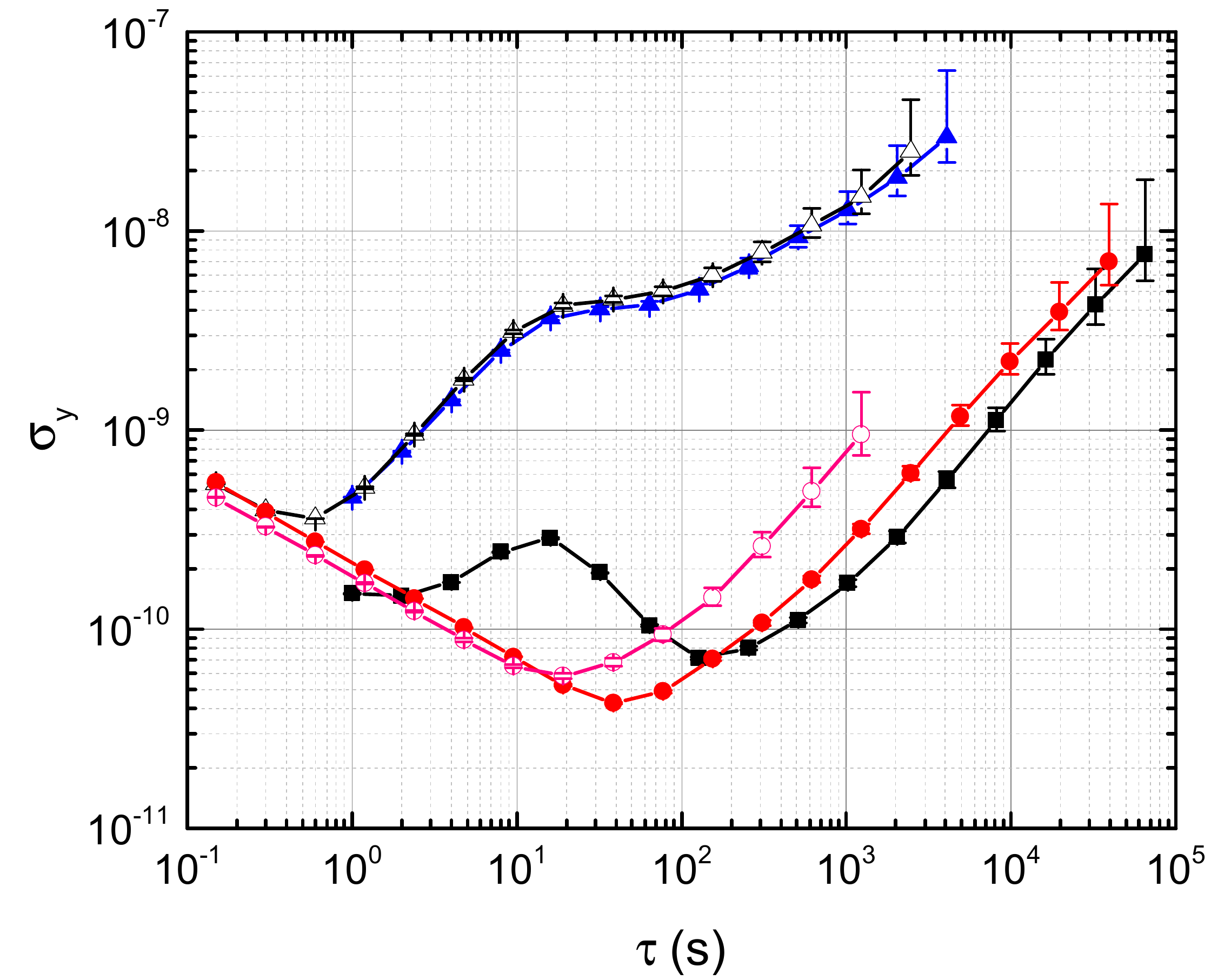}%
\caption{Comparison of calculated Allan deviations: ultra-stable laser [measurement performed by the wavelength meter when it was not isolated (pink3 open circles) and when it was encompassed in a thermally isolated box (red circles)], free-running LUT (measurement performed by the wavelength meter; black open triangles), free-running laser under test (LUT) (measurement performed by the frequency counter; blue triangles) and LUT stabilized by the wavelength meter (measurement performed by the frequency counter; black squares).}.%
\label{fig:stab}%
\end{figure}

Once the required power was detected by the wavelength meter, we were able to estimate the wavelength meter relative frequency instability. To do so, we have computed the Allan deviation of the measured wavelength (771 nm) when using the ultra-stable laser frequency-doubled signal as an input. The ultra-stable laser frequency instability has been checked to be well below $1{\times}10^{-11}$ from 1 s to 10000 s during the wavelength meter characterization. Since the ultra-stable laser frequency instability is far below the wavelength meter instability, the resulting Allan deviation is a good estimate of the wavelength meter relative frequency stability. The result is shown in Fig. \ref{fig:stab} (open circles). The presented result shows a relative frequency instability of $2{\times}10^{-10} \ \tau^{-1/2}$ from 0.15 s to 10 s and a long-term frequency drift of $8{\times}10^{-12} \ \tau$ (about 27 MHz/day).

As we have suspected a thermal sensitivity of the wavelength meter to the variations of the room temperature, we have encompassed the wavelength meter in a thermally isolated box and the measurement was repeated. The temperature fluctuations in the box are reduced by an order of magnitude at $\tau=100$s as compared to the room temperature fluctuations. The Allan deviation of the new measurement is shown in Fig. \ref{fig:stab} (red circles). As one can see, we were able to reduce the wavelength meter frequency drift down to about $3{\times}10^{-13}\ \tau$ (about 10 MHz/day). This reduction in the frequency drift is also perceptible in the recorded optical frequencies in the time domain for slightly more than three hours only. As a result, one can conclude that the stability of laser sources featuring a frequency instability above the red curve presented in Fig. \ref{fig:stab} can be accurately characterized by this wavelength meter. Moreover, the obtained relative frequency instability (Fig. \ref{fig:stab}, red circles) indicates a frequency resolution far below the instrument technical specification given by the manufacturer.

The frequency instability of the wavelength meter is in fair agreement with previously published results. Kobtsev \emph{et al.} measured frequency drifts much higher than what we obtained without the thermal insulation box but used a previous generation of wavelength meter \cite{Kobtsev2007}. The drift measured in \cite{Pyka2014} is of the same order of magnitude than what we obtained without the insulation box. As for the short-term fluctuations, the only value known to us is published in \cite{Pyka2014} and corresponds to a relative frequency instability of about $2.2{\times}10^{-9}$. This is an order of magnitude higher than our findings, but the specific conditions of the measurement (especially the optical power sent to the wavelength meter) may have impacted its frequency instability.

\section{Laser frequency stabilization by the wavelength meter}

Since we now know the limits of the wavelength meter in terms of frequency instability, we have performed different experimental tests to explore the wavelength meter performance in terms of frequency-locking. For this purpose, we have used a dedicated characterization setup including a second telecom laser (a second Koheras AdjustiK$^{\mathrm{TM}}$ E15 from NKT Photonics).

\subsection{Frequency-lock characterization setup}

\begin{figure}%
\centering
\includegraphics[width=\columnwidth]{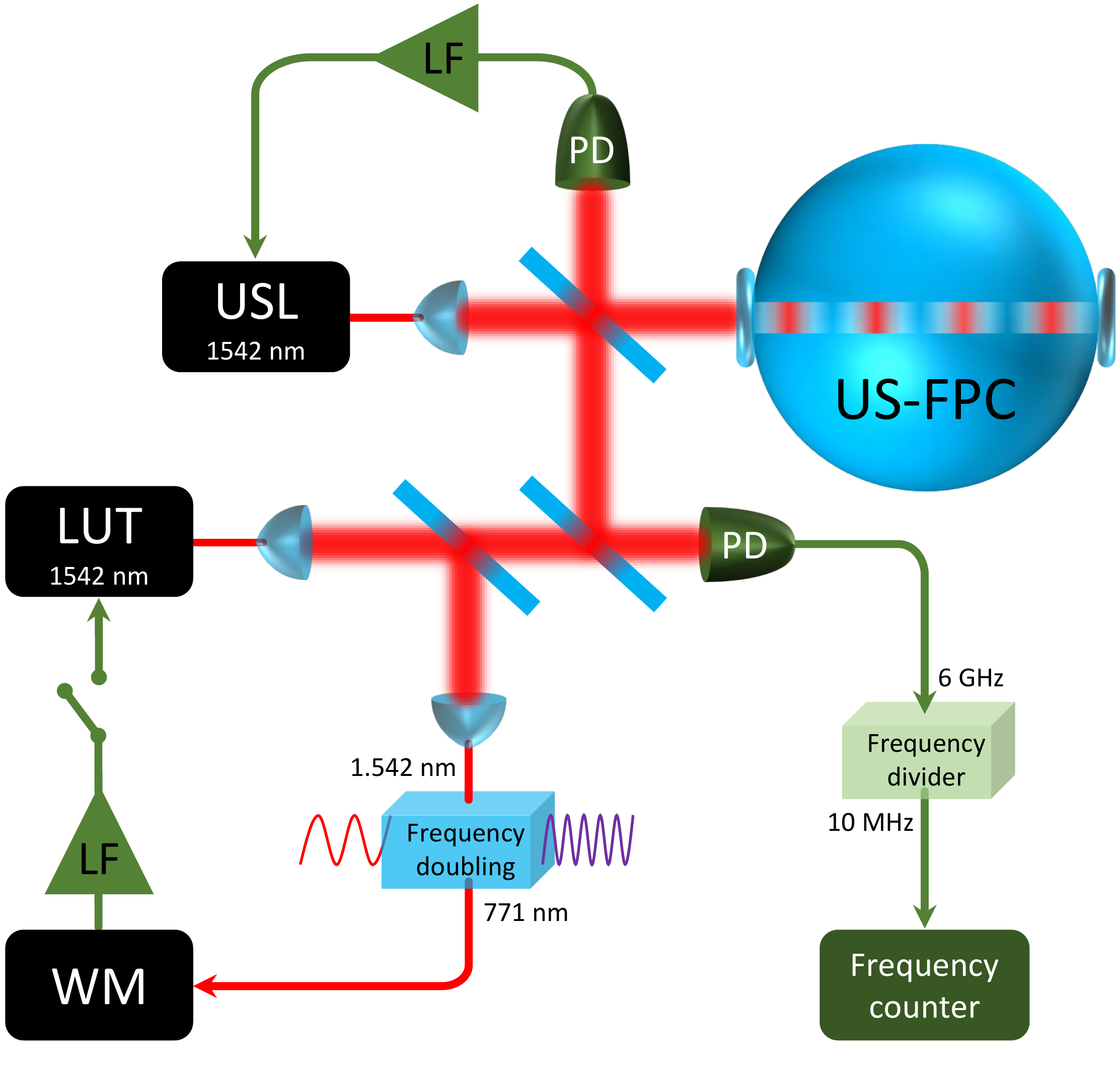}%
\caption{Frequency-lock characterization setup. The laser under test (LUT) is frequency-doubled so that its frequency can be measured and corrected by the wavelength meter. An optical beatnote between the LUT and an ultra-stable laser is detected by a photodiode and divided down to be measured by a frequency counter for characterization. PD: photodiode; LF: loop filter.}.%
\label{fig:setup_lock}%
\end{figure}

We have used the experimental setup depicted in Fig. \ref{fig:setup_lock} in order to characterize the wavelength meter frequency stabilization performance. In this setup, the aforementioned ultra-stable laser was used with a second telecom laser considered as the LUT. First, a beam splitter was used to extract a fraction of the LUT beam ($\approx$20 mW) that was then sent to the PPLN frequency doubling crystal. The frequency-doubled signal of the LUT was then sent to the wavelength meter to be measured and to generate a control signal in order to stabilize the LUT frequency. On the other hand, the 1542 nm fraction of the LUT beam was combined with a fraction of the ultra-stable laser beam. The optical beatnote was detected by a fast photodiode featuring a 10 GHz bandwidth (a DSC30S photodiode from Discovery Semiconductors). We have then manually tuned the LUT optical frequency to be at approximately 6 GHz from the ultra-stable laser frequency. To characterize the frequency instability of the 6 GHz beatnote at the photodiode output, we divided the beatnote signal frequency down to 10 MHz. The resulting signal was finally sent to a frequency counter referenced to a hydrogen maser.

\subsection{Free-running laser instability measurements}

Firstly, we have computed the Allan deviation of the measured wavelength (771 nm) when using the free-running LUT frequency-doubled signal as an input for the wavelength meter. We have also computed the Allan deviation of the beatnote between the ultra-stable laser and the free-running LUT, measured by the frequency counter. Both results are respectively depicted in Fig. \ref{fig:stab} (blue trinalges, black open triangles). These results, and the corresponding measured frequencies using the wavelength meter and the frequency counter, show a very good agreement between the measurements performed using the wavelength meter and those performed using the optical beatnote signal. This confirms the very good reliability of the wavelength meter as a frequency instability characterization instrument. 

\begin{figure}%
\centering
\includegraphics[width=\columnwidth]{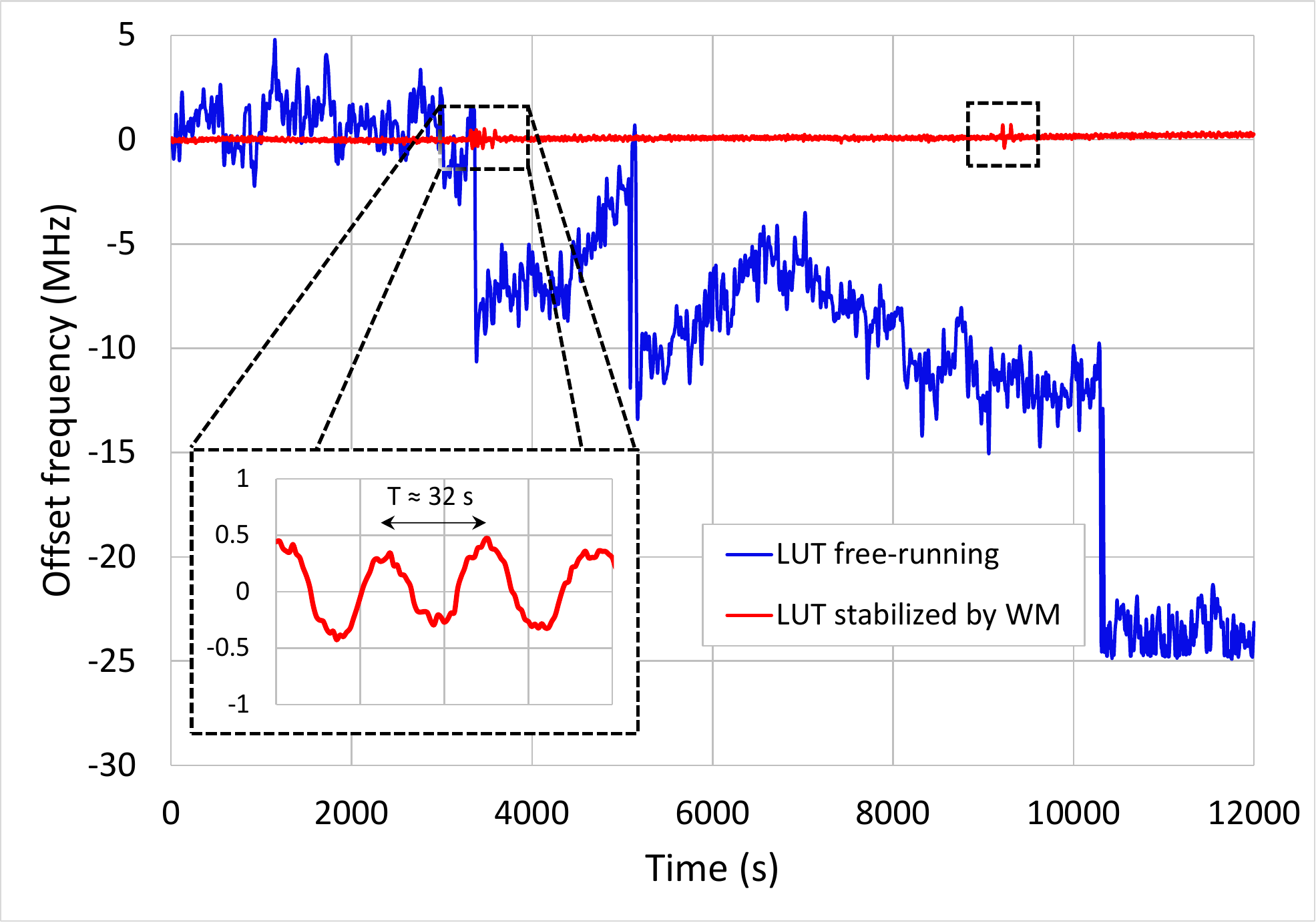}%
\caption{Temporal variations of the LUT frequency measured by the frequency counter: free-running LUT (dark-blue curve) and LUT stabilized by the wavelength meter (black curve).}.%
\label{fig:temp}%
\end{figure}

\subsection{Laser frequency-locking by the wavelength meter}

A frequency-lock of the LUT was performed using the wavelength meter. For this purpose, we had to first set the appropriate gain for the wavelength meter digital loop filter and define the reference wavelength to which the LUT would be stabilized. The control signal acted on the LUT piezoelectric transducer. Once the LUT was locked, we have measured its frequency instability using the optical beatnote scheme and the frequency counter. The calculated Allan deviation is represented in Fig. \ref{fig:stab} (black squares) and the corresponding measured frequencies are shown in black in FIG. \ref{fig:temp}. These results, and more clearly Fig. \ref{fig:stab} results, reveals that the LUT had a lower short-term frequency instability than the wavelength meter ($1.5{\times} 10^{-10}$ at 1s vs $2{\times}10^{-10}$ at 1s for the wavelength meter). This has also been confirmed by comparing the phase noise spectra of the optical beatnote when the LUT was locked and when it was in free-running regime. Additionally, from Fig. \ref{fig:stab} results, we can see that the locked LUT had a lower frequency drift ($1.4{\times}10^{-13} \ \tau$; about 2.3 MHz/day at 1542 nm) than what was previously measured for the wavelength meter ($3{\times}10^{-13} \ \tau$; about 10 MHz/day). This difference in the frequency drift was most probably due to room temperature variations which differently affected the two measurements as they were performed at different times. On the other hand, the peak shown at $\tau$ = 16 s was due to oscillations in the LUT frequency resulting from small frequency jumps in the LUT (see the inset figure in Fig. \ref{fig:temp}) that were not fully rejected due to a lack of gain in the control loop. The origin of the frequency deviations in our LUT is still unidentified.

All things considered, we can see that a frequency instability below $3{\times}10^{-10}$ from 1 to 2000s was achieved for the LUT stabilized by the wavelength meter. This confirms that such instruments have very good measurement and frequency-control performances. In particular, this is slightly lower than the results obtained with discharge lamps \cite{Streed2008}. These performances are particularly interesting for several applications in the time and frequency domain and in fundamental physics experiments (ex. ions trapping, atoms cooling, etc.).

\section{Conclusion}

We have characterized a commercial wavelength meter that exhibits a frequency instability $\sigma_y(\tau)= 2{\times}10^{-10} \ \tau^{-1/2}$ from 0.15 s to 30 s, and a long-term frequency drift of $3 {\times} 10^{-13} \ \tau$ (about 10 MHz/day). This indicates a frequency resolution far below the apparatus technical specification. We demonstrate laser frequency locking to the wavelength meter with a residual frequency instability below $3{\times}10^{-10}$ for integration times between 1 s and 2000 s, low enough to be used in a laser cooling experiment.

The frequency drift of the WS7 is lower than typical low-finesse cavities, and is sufficiently low for operating a laser cooling setup over the course of a day. Locking or comparison to an atomic reference will still be necessary for continuous operation on longer timescales, like is typically required in atomic clock setups that require long integration times. This drift could also be lowered by actively locking the wavelength meter temperature.

In conclusion, Fizeau wavelength meters have proven to be versatile experimental tools allowing both characterization and frequency stabilization of narrow-linewidth lasers. In addition, one can use fibered MEMS switch to stabilize several lasers using the same wavelength meter, relieving from the need of a complex frequency stabilization optical bench. In this manuscript, we have fully characterize the frequency resolution and stability of such an apparatus, providing guidelines and bounds to the performances they can achieve.

\section{Acknowledgements}

The authors would like to thank Rodolphe Boudot and Enrico Rubiola for fruitful discussions. We gratefuly acknowledge the help of Thibaut Sylvestre, who lended us the optical frequency doubling cristal.

We thank the  Agence Nationale de la Recherche (ANR) for financial support under grant ANR-14-CE26-0031-01-MITICC. This work is also funded by the ANR Programme d'Investissement d'Avenir (PIA) under the First-TF network, and by grants from the R\'egion Franche Comt\'e.

\bibliographystyle{osajnl}
\bibliography{bib_ws7}
%

\end{document}